\documentclass[titlepage,oneside,a4paper,reqno]{amsart} 
\usepackage{amsmath}
\usepackage{xspace}
\usepackage[mathscr]{eucal}
\newtheorem*{proposition}{Proposition}
\newcommand{\mat}[3]{\ensuremath{
																								\left \langle  \vphantom{#2 #3}   #1   
                        \right|    					\, #2\,   
                        \left|    \vphantom{#2 #1} #3   
                        \right \rangle
                        								}
                     }
\newcommand{\bmat}[3]{\ensuremath{
																								\bigl \langle     #1   \bigr|    \, #2\,   \bigl|     #3   \bigr \rangle
                        									}
                     }
\newcommand{\ket}[1]{\ensuremath{		\left| #1 \right> 
																																			  }
																									}
\newcommand{\overlap}[2]{\ensuremath{ 
																								\left \langle    #1 \vphantom{#2 } \,
                        \right| \left.   #2 \vphantom{#1}
                        \right \rangle
                        									}
                     }

\newcommand{\comm}[2]{\ensuremath{  \left[ #1, #2 \right] }}
\newcommand{\bcomm}[2]{\ensuremath{							\bigl[ #1 , #2 \bigr]							}}
\newcommand{\pt}{   \ensuremath{     \phi_{\mathrm{ap}}     }}
\newcommand{\xOp}{ \ensuremath{  \hat{x}  }}
\newcommand{\pOp}{ \ensuremath{  \hat{p}  }}
\newcommand{\xiOp}{ \ensuremath{  \hat{x}_{\mathrm{i}}  }\xspace}
\newcommand{\piOp}{ \ensuremath{  \hat{p}_{\mathrm{i}}  }\xspace}
\newcommand{\xfOp}{ \ensuremath{  \hat{x}_{\mathrm{f}}  }\xspace}
\newcommand{\pfOp}{ \ensuremath{  \hat{p}_{\mathrm{f}}  }\xspace}
\newcommand{\MxOp}{  \ensuremath{   \hat{\mu}_{\mathrm{X}}    } }
\newcommand{\MpOp}{  \ensuremath{   \hat{\mu}_{\mathrm{P}}    } }

\newcommand{\MxfOp}{  \ensuremath{   \hat{\mu}_{\mathrm{Xf}}    } }
\newcommand{\MpfOp}{  \ensuremath{   \hat{\mu}_{\mathrm{Pf}}    } }
\newcommand{\Mx}{   \ensuremath{   \mu_{\mathrm{X}}  } }
\newcommand{\Mp}{   \ensuremath{   \mu_{\mathrm{P}}  } }
\newcommand{\Mxf}{   \ensuremath{   \mu_{\mathrm{Xf}}  } }
\newcommand{\Mpf}{   \ensuremath{   \mu_{\mathrm{Pf}}  } }
\newcommand{\PxOp}{  \ensuremath{   \hat{\pi}_{\mathrm{X}}    } }
\newcommand{\PpOp}{  \ensuremath{   \hat{\pi}_{\mathrm{P}}    } }

\newcommand{\PxfOp}{  \ensuremath{   \hat{\pi}_{\mathrm{Xf}}    } }
\newcommand{\PpfOp}{  \ensuremath{   \hat{\pi}_{\mathrm{Pf}}    } }
\newcommand{\Exi}{\ensuremath{       \hat{\epsilon}_{\mathrm{Xi}}       }\xspace}
\newcommand{\Epi}{\ensuremath{       \hat{\epsilon}_{\mathrm{Pi}}       }\xspace}
\newcommand{\Exf}{\ensuremath{       \hat{\epsilon}_{\mathrm{Xf}}       }\xspace}
\newcommand{\Epf}{\ensuremath{       \hat{\epsilon}_{\mathrm{Pf}}       }\xspace}
\newcommand{\Dx}{\ensuremath{       \hat{\delta}_{\mathrm{X}}       }\xspace}
\newcommand{\Dp}{\ensuremath{       \hat{\delta}_{\mathrm{P}}       }\xspace}
\newcommand{\RErr}{\ensuremath{     \Delta_{\mathrm{ei}}  }\xspace}
\newcommand{\PErr}{\ensuremath{     \Delta_{\mathrm{ef}}  }\xspace}
\newcommand{\Dist}{\ensuremath{     \Delta_{\mathrm{d}}  }\xspace}
\begin{document}
\title{The Concept of Experimental Accuracy and Simultaneous Measurements of Position and Momentum} 
\author{D.M.Appleby}
\begin{titlepage}
\begin{center}
\bfseries
THE CONCEPT OF EXPERIMENTAL ACCURACY AND SIMULTANEOUS MEASUREMENTS OF POSITION AND MOMENTUM
\end{center}
\vspace{1 cm}
\begin{center}
D.M.APPLEBY\footnote{Dept. of Physics, Queen Mary and
		Westfield College, 327 Mile End Rd., London E1 4NS, U.K.

E-mail Address:  D.M.Appleby@qmw.ac.uk}
\end{center}
\vspace{2.5 cm}
\begin{center}
\textbf{Abstract}\\
\vspace{0.35 cm}
\parbox{10.5 cm }{
The concept of experimental accuracy is investigated in the context of the 
unbiased joint measurement processes defined by Arthurs and Kelly.  
A distinction is made
between the errors of retrodiction and prediction.  Four error-disturbance relationships are
derived, analogous to the single error-disturbance relationship derived by Braginsky and
Khalili in the context of single measurements of position only.
A retrodictive and a predictive error-error relationship are also derived.  The connection between
these relationships and the extended Uncertainty Principle of Arthurs and Kelly is discussed.  The
similarities and differences between the quantum mechanical and classical concepts of experimental
accuracy are explored.  It is argued that  these relationships  provide grounds for questioning
Uffink's conclusion, that the concept of a simultaneous measurement of non-commuting observables is
not fruitful.}
\end{center}
\end{titlepage}
\section{Introduction}
\label{sec:  intro}
Notwithstanding the fundamental importance of the uncertainty principle there is still,
as Hilgevoord and Uffink (1990) have remarked, a great deal of discussion about
what it  actually says.  The purpose of this paper is to add a few additional points to the 
discussion.  We are particularly concerned with the idea 
that the uncertainty principle represents a
constraint on the accuracy achievable
in a simultaneous measurement of position and momentum.

The form of the uncertainty principle given in most modern textbooks is the inequality
\begin{equation}
   \Delta x  \Delta p  \ge \frac{\hbar}{2}
\label{eq:  UncertRel}
\end{equation}
where the quantities $\Delta x$, 
$\Delta p$ are defined in terms of the state of the system $\ket{\psi}$ by
\begin{equation}
\begin{split}
    \Delta x  
& = \sqrt{   \bmat{\psi}{  \hat{x}^2 }{ \psi} - \bmat{\psi}{  \hat{x} }{ \psi}^2}
\\
    \Delta p  
& = \sqrt{   \bmat{\psi}{  \hat{p}^2 }{ \psi} - \bmat{\psi}{  \hat{p} }{ \psi}^2}
\end{split}
\label{eq:  UncertsDef}
\end{equation}
The first general proof of inequality~(\ref{eq:  UncertRel}) was actually
given by Kennard (1927), not Heisenberg.  We will accordingly refer to this form
of the uncertainty principle as Kennard's inequality.

The proof of Kennard's inequality is based on
the fact that   $\overlap{p}{\psi}$ is the Fourier transform of $\overlap{x}{\psi}$.
In his original paper, and again in his Chicago lectures, Heisenberg (1927, 1930) also gave 
another, quite different argument involving a 
$\gamma$-ray microscope.  On the basis
of this argument he interpreted $\Delta x$ and $\Delta p$ as experimental
\emph{errors} or \emph{inaccuracies}.  He thereby suggested that the uncertainty
principle should be understood to mean, in the words of Bohm (1951),
\begin{quote}
  If a measurement of position is made with accuracy $\Delta x$, and if a measurement 
  of momentum is made \emph{simultaneously} with accuracy $\Delta p$, then the product of the 
  two errors can never be smaller than a number of order $\hbar$.
\end{quote}
(Bohm's emphasis).
Heisenberg himself did not state the matter quite so plainly; however, one has the impression
that he would have concurred with the above statement of Bohm's
had it been put to him.  
The question arises:  
is this a valid interpretation of Kennard's inequality?  
The question has been discussed by
Ballentine (1970), 
W\'{o}dkiewicz (1987),
Hilgevoord and Uffink (1990),
Raymer (1994) and de Muynck \emph{et al} (1994).  We will here confine
ourselves to remarking that the quantities $\Delta x$ and $\Delta p$ defined
by Eqs.~(\ref{eq:  UncertsDef}) cannot be interpreted as experimental errors in
anything like the normal sense of the word because they only depend on the
state $\ket{\psi}$.  They are thus \emph{intrinsic} properties of the isolated 
system.  An experimental error, by contrast, ought to depend on the state 
of the measuring apparatus, as well as the state of the system.  In other words,
it should partly depend on quantities which are \emph{extrinsic} to the system.

Suppose, that in Heisenberg's microscope \emph{gedanken} experiment, one were to 
make the microscope go out of focus.  This should have the effect of increasing 
the error
in the measurement of $x$.  But it will have no effect on the quantity 
$\Delta x$, since this only depends on the initial state of the particle whose
position is being measured.

These considerations do not mean that the statement of Bohm's quoted above
is incorrect.  They do, however, mean that it is not a consequence of 
the inequality
proved by Kennard.   Rather, it represents (if true) an independent physical
principle.  For the sake of distinctness let us give it a name:  the error principle.

The problem we now face is, that whereas there exists a rigorous mathematical
proof of Kennard's inequality, the status of the error principle is much 
more ambiguous.   Indeed, the very meaning of the concepts involved---the concept of
a simultaneous measurement of position and momentum, and the concept of 
experimental accuracy---continues to be the subject of discussion.

One approach to the problem is that
based on the concept of a ``fuzzy,'' or ``stochastic'' measurement, due
to Prugove\v{c}ki (1984),  Holevo (1982), Busch and Lahti (1984),
Martens and de Muynck (1992), de Muynck \emph{et al} (1994) and others
[for additional references see the works just cited and Uffink (1994)].
Uffink (1994) has identified a number of objections to this approach.
His conclusion is  ``that the claim that within this formalism a joint unsharp
measurement of position and momentum \dots is possible is false''.  Moreover, he
doubts whether matters could be remedied by adopting a different approach.
He considers that ``the formalism of quantum theory, as presented by von Neumann, 
simply has no room for a description of a joint measurement of position and 
momentum at all''---not even a less than perfectly accurate joint measurement.

We acknowledge the force of Uffink's arguments.  Nevertheless, we are unwilling 
to accept his analysis as the last word on the subject.
In the first place, ordinary laboratory practice depends on the assumption that
it is possible to make simultaneous, imperfectly accurate 
determinations of the position and momentum of macroscopic
objects.  If it is true that quantum mechanics does not allow for the existence
of such measurements, then one of two things would seem to follow: 
either normal laboratory practice
is based on a misconception, in which case much of the evidential basis for modern physics
(including quantum mechanics) would simply collapse; or
else quantum mechanics does not apply on the macroscopic scale.  In short, Uffink's
conclusion has some fairly momentous consequences.  This is not, of course,
a reason for rejecting Uffink's conclusion.  It is, however, a reason for re-examining
the question, to see if there is some way of avoiding his conclusion.

In the second place, a number of authors 
(Arthurs and Kelly, 1965; Braunstein \emph{et al}, 1991; Stenholm, 1992; 
Leonhardt and Paul, 1993;
T\"{o}rma \emph{et al}, 1995) have described several specific processes
which might be described (and which they do describe) as simultaneous
measurements of position and momentum.  Their work is logically independent 
of the work criticised by Uffink, and it is therefore not open to the same
objections.  Indeed, Uffink explicitly states that he does not mean to impugn
the approach of these authors (although he does question whether it is 
``fruitful'' to interpret the processes they describe as simultaneous measurements
of non-commuting observables).

Within the context of their approach Arthurs and Kelly (1965) have derived
an ``extended'' or ``generalised'' uncertainty principle (also see
W\'{o}dkiewicz, 1987; Arthurs and Goodman, 1988;  Raymer, 1994; Leonhardt and Paul, 1995).
Let $\Delta \Mxf$ (respectively $\Delta \Mpf$) be the standard deviation for the outcome of
the measurement of $\hat{x}$ (respectively $\hat{p}$).  Then, subject to certain restrictive
assumptions regarding the nature of the measurement process, Arthurs and Kelly show
\begin{equation}
    \Delta \Mxf \, \Delta \Mpf  \ge \hbar
\label{eq:  AKExtUncertPrinc}
\end{equation}
where we have employed a different notation from that of Arthurs and Kelly (the reasons
for this notation will become clear in the next section).

The quantities $\Delta \Mxf$ and $\Delta \Mpf$ are not interpretable as experimental
errors.  However, they do depend on the initial state of the apparatus, as well as the
initial state of the system.  Moreover, the increase in the lower bound set by
inequality~(\ref{eq:  AKExtUncertPrinc}) as compared with 
Kennard's inequality can be taken as a quantitative indication of
the noise introduced by the measurement.  So, although 
this relation cannot be regarded as a quantitative expression of the 
error principle (the statement of Bohm's quoted above), it may at least be regarded
as a step in that direction.

Another relation relevant to our problem is the one derived by
Braginsky and Khalili~(1992), in the context of  single measurements
of position only.  Braginsky and Khalili define a quantity $\Delta x_{\mathrm{measure}}$,
representing the error in the measurement of $\hat{x}$, and a quantity
$\Delta p_{\mathrm{perturbation}}$, representing the disturbance of the
conjugate quantity $\hat{p}$; and they show
\begin{equation}
    \Delta x_{\mathrm{measure}} \, \Delta p_{\mathrm{perturbation}}  \ge \frac{\hbar}{2}
\label{eq:  BKExtUncertPrinc}
\end{equation}
provided that the measurement is of the special kind which they describe as linear.  
Their inequality does not refer to simultaneous measurements of position \emph{and}
momentum, and only one of the two quantities on the left hand side is interpretable
as an experimental error.  However, its existence encourages us to believe that a similar
approach might prove fruitful in the problem of interest here.

The purpose of this paper is to combine and to develop the approaches of Arthurs and Kelly
on the one hand,
and of Braginsky and Khalili on the other, in an attempt to find a precise, quantitative expression
of the error principle as stated by Bohm in the passage quoted above.  

The result of our 
analysis is to show that there are in fact two different error principles, corresponding
to the predictive and retrodictive aspects of a measurement process as discussed by
Hilgevoord and Uffink (1990) (also see Prugove\v{c}ki, 1973, 1975). In addition, we derive
four error-disturbance relationships (in place of the single
relationship derived by Braginsky and Khalili).  

The six inequalities which we derive in the following sections, together with
Kennard's inequality, gives a total of
seven inequalities.  If our analysis is correct all of these inequalities
are needed to capture the full intuitive content of Heisenberg's
original paper.

\section{The Arthurs-Kelly Process}
We begin by considering a specific example of a simultaneous measurement process; 
namely the process described by Arthurs and Kelly (1965) (also see 
Braunstein \emph{et al}, 1991; Stenholm, 1992).  Suppose that we have a
system interacting with a measuring apparatus, or meter.   
The system has one degree of freedom, with 
position $\hat{x}$ and conjugate momentum $\hat{p}$.  The measuring apparatus has two
degrees of freedom, comprising two pointer observables
$\MxOp$, $\MpOp$ with conjugate momenta $\PxOp$, $\PpOp$.    The pointer observables
$\MxOp$, $\MpOp$ give the result of the measurement.  We have
the commutation relations
\begin{equation*}
 \comm{\hat{x}}{\hat{p}} = \comm{\MxOp}{\PxOp} = \comm{\MpOp}{\PpOp} = i \hbar
\end{equation*}
these being the only non-vanishing commutators between the six operators
$\hat{x}$, $\hat{p}$, $\MxOp$,  $\PxOp$,  $\MpOp$,  $\PpOp$.

The unitary evolution operator describing the measurement interaction is
\begin{equation*}
  \hat{U}  =  e^{ - \frac{i}{\hbar} ( \PpOp \hat{p}  +  \PxOp \hat{x} )}
\end{equation*}

Suppose that the system $+$ apparatus composite is initially in the product
state \ket{\psi \otimes \pt}, where \ket{\psi} is the initial state of the system,
and \ket{\pt} is the initial state of the apparatus.  
The probability distribution 
for the result of the measurement is then given by
\begin{equation*}
     \rho ( \Mx, \Mp )  
=    \int dx \, \bigl|\bmat{ x , \Mx, \Mp }{\hat{U}}{ \psi \otimes \pt}\bigr|^2
\end{equation*}

In order to describe the experimental errors, and the disturbance of the system by 
the measurement process, it is convenient to switch to the Heisenberg picture.
Let $\hat{\mathcal{O}}$ be any of the operators
$\xOp$, $\pOp$, $\MxOp$,  $\PxOp$,  $\MpOp$,  $\PpOp$.   We then define the
initial Heisenberg picture operator $\hat{\mathcal{O}}_{\mathrm{i}}$ and 
final Heisenberg picture operator $\hat{\mathcal{O}}_{\mathrm{f}}$ by
\begin{equation*}
  \begin{split}
       \hat{\mathcal{O}}_{\mathrm{i}} & = \hat{\mathcal{O}} \\
       \hat{\mathcal{O}}_{\mathrm{f}} & = \hat{U}^{\dagger} \hat{\mathcal{O}} \hat{U}
  \end{split}
\end{equation*}
It is readily found
\begin{equation}
  \begin{aligned}
      \xfOp & = \hat{U}^{\dagger} \xOp \hat{U} & & = \xOp + \PpOp \\
      \pfOp & = \hat{U}^{\dagger} \pOp \hat{U} & & = \pOp - \PxOp \\
      \MxfOp & = \hat{U}^{\dagger} \MxOp \hat{U} & & = \MxOp + \xOp + \tfrac{1}{2} \PpOp \\
      \MpfOp & = \hat{U}^{\dagger} \MpOp \hat{U} & & = \MpOp + \pOp - \tfrac{1}{2} \PxOp \\
      \PxfOp & = \hat{U}^{\dagger} \PxOp \hat{U} & & = \PxOp  \\
      \PpfOp & = \hat{U}^{\dagger} \PpOp \hat{U} & & = \PpOp  \\
  \end{aligned}
\label{eq:  AKHeisFinOps}
\end{equation}
We now define the retrodictive error operators
\begin{equation}
  \begin{split}
       \Exi  & = \MxfOp - \xiOp \\
       \Epi  & = \MpfOp - \piOp 
  \end{split}
\label{eq:  RetErrOps}
\end{equation}
the predictive error operators
\begin{equation}
  \begin{split}
       \Exf  & = \MxfOp - \xfOp \\
       \Epf  & = \MpfOp - \pfOp 
  \end{split}
\label{eq:  PreErrOps}
\end{equation}
and the disturbance operators
\begin{equation}
  \begin{split}
       \Dx  & = \xfOp - \xiOp \\
       \Dp  & = \pfOp - \piOp
  \end{split}
\label{eq:  DisOps}
\end{equation}
The motivation for these definitions will be clearest if we think, for a moment,
in classical terms.  In that case \Exi, \Epi  give the difference
between the final pointer positions and the initial system  observables
$\xiOp$, $\piOp$.  In other words they tell us how accurately the result of the measurement
reflects the initial state of the system, \emph{before} the measurement was carried out, which
is why we refer to them as retrodictive error operators.  On the 
other hand \Exf, \Epf  give the difference between the final pointer 
positions and the final system observables $\xfOp$, $\pfOp$.  They therefore tell us
how accurately the result of the measurement reflects the final state of the system,
\emph{after} the measurement has been completed, which is why we refer to 
them as predictive error operators.  Lastly, 
\Dx, \Dp give the difference between the final system observables
$\xfOp$, $\pfOp$  and the initial system observables $\xiOp$, $\piOp$.  They therefore
describe the disturbance of the system by the measurement process.

Of course, we are actually talking about quantum mechanics, not classical mechanics.  Our
definitions therefore raise some important conceptual questions.  We  do
not wish to minimise these questions.   We do, however, wish to defer discussing
them until after we have derived some quantitative formulae.  
It is to be observed,
that whatever the precise conceptual, or philosophical status of the quantities just
introduced, they are well-defined mathematically.

In order to obtain  numerical indications of the accuracy and 
disturbance we take the rms values of the operators just defined.  We thus
have, the  rms errors of retrodiction
\begin{equation}
\begin{split}
    \RErr x  & = \sqrt{ \mat{ \psi \otimes \pt}{\Exi^2}{ \psi \otimes \pt} } \\
    \RErr p  & = \sqrt{ \mat{ \psi \otimes \pt}{\Epi^2}{ \psi \otimes \pt} }
\end{split}
\label{eq:  AKrmsRerr}
\end{equation}
the  rms errors of prediction
\begin{equation}
\begin{split}
    \PErr x  & = \sqrt{ \mat{ \psi \otimes \pt}{\Exf^2}{ \psi \otimes \pt} } \\
    \PErr p  & = \sqrt{ \mat{ \psi \otimes \pt}{\Epf^2}{ \psi \otimes \pt} }
\end{split}
\label{eq:  AKrmsPerr}
\end{equation}
and the  rms disturbances
\begin{equation}
\begin{split}
    \Dist x  & = \sqrt{ \bmat{ \psi \otimes \pt}{\Dx^2}{ \psi \otimes \pt} } \\
    \Dist p  & = \sqrt{ \bmat{ \psi \otimes \pt}{\Dp^2}{ \psi \otimes \pt} }
\end{split}
\label{eq:  AKrmsDis}
\end{equation}

The above definitions  apply to any measurement process.  Let
us now specialise to the case of the Arthurs-Kelly process.   
Inserting~(\ref{eq:  AKHeisFinOps}) in 
the defining equations~(\ref{eq:  RetErrOps}--\ref{eq:  DisOps}) gives
\begin{equation}
\begin{aligned}
   \Exi & = \MxOp + \tfrac{1}{2} \PpOp & \hspace{0.5 in} 
   \Exf & = \MxOp - \tfrac{1}{2} \PpOp & \hspace{0.5 in} 
   \Dx  & = \phantom{-} \PpOp 
\\
   \Epi & = \MpOp - \tfrac{1}{2} \PxOp & \hspace{0.5 in}
   \Epf & = \MpOp + \tfrac{1}{2} \PxOp & \hspace{0.5 in} 
   \Dp  & = - \PxOp 
\end{aligned}
\label{eq:  AKErrDisOps}
\end{equation}
It is to be observed that the error and disturbance operators only depend on
the pointer positions and momenta.  It follows, that the rms errors and disturbances
as defined by equations~(\ref{eq:  AKrmsRerr}--\ref{eq:  AKrmsDis}) are independent
of the initial system state.  This is, of course, a peculiarity of the Arthurs-Kelly
process.  We do not expect it to be true generally.

Using equations~(\ref{eq:  AKErrDisOps}) we find
\begin{equation}
\begin{aligned}
  \bcomm{\Exi}{\Epi} & = - i \hbar & \hspace{0.5 in}
  \bcomm{\Exi}{\Dp}  & = - i \hbar & \hspace{0.5 in}
  \bcomm{\Dx}{\Epi}  & = - i \hbar & \hspace{0.5 in}
\\
  \bcomm{\Exf}{\Epf} & = \phantom{-} i \hbar & \hspace{0.5 in}
  \bcomm{\Exf}{\Dp}  & = - i \hbar & \hspace{0.5 in}
  \bcomm{\Dx}{\Epf}  & = - i \hbar & \hspace{0.5 in}
\end{aligned}
\label{eq:  AKErrDisComRels}
\end{equation}
these being the only non-vanishing commutation relationships between members
of the set \Exi, \Epi, \Exf, \Epf, \Dx, \Dp.  Taking this result in conjunction 
with the defining equations~(\ref{eq:  AKrmsRerr}--\ref{eq:  AKrmsDis}) we deduce,
a retrodictive error relationship
\begin{equation}
   \RErr x \, \RErr p  \ge \frac{\hbar}{2}
\label{eq:  AKRetErrRel}
\end{equation}
a predictive error relationship
\begin{equation}
   \PErr x \, \PErr p  \ge \frac{\hbar}{2}
\label{eq:  AKPreErrRel}
\end{equation}
and four error-disturbance relationships
\begin{equation}
\begin{aligned}
    \RErr x  \,  \Dist p & \ge \frac{\hbar}{2} & \hspace{0.5 in}
    \RErr p \,  \Dist x & \ge \frac{\hbar}{2}   \hspace{0.5 in}
\\
    \PErr x  \,  \Dist p & \ge \frac{\hbar}{2} & \hspace{0.5 in}
    \PErr p \,  \Dist x & \ge \frac{\hbar}{2}   \hspace{0.5 in}
\end{aligned}
\label{eq:  AKErrDisRels}
\end{equation}
Equations~(\ref{eq:  AKRetErrRel}) and~(\ref{eq:  AKPreErrRel}) together constitute
a quantitative expression of the semi-intuitive error principle, as stated by
Bohm~(1951) in the passage quoted earlier.
Equations~(\ref{eq:  AKErrDisRels}) provide a quantitative expression of the principle,
that an increased degree of accuracy in the measurement of one observable can only be achieved
at the expense of an increased degree of disturbance in the 
canonically conjugate observable.

The reason one needs two inequalities to capture the full content 
of the error principle is the fact that one has to distinguish the errors
of prediction from the errors of retrodiction.  In classical physics it is not usually
necessary to emphasise this distinction.  This is because, in classical physics, the 
disturbance of the system by the measurement can, in principle, be made negligible.  In 
quantum mechanics, however, the back-reaction of the apparatus on the system is very
important.
As a result, the distinction between the two kinds of error is also essential.  In fact,
it is an immediate consequence of the definitions that
\begin{equation*}
\begin{split}
   \Dx & = \Exi - \Exf \\
   \Dp & = \Epi - \Epf
\end{split}
\end{equation*}
It follows, that if the disturbances cannot be assumed to be negligible, then
neither can the difference between the retrodictive and predictive errors.

The reason that there are four error-disturbance relations in our analysis, but only one
in the analysis of Braginsky and Khalili is; firstly, that Braginsky and Khalili
do not consider simultaneous measurements of $\hat{x}$ and $\hat{p}$; and
secondly, that they only consider  the error of retrodiction (as we have termed it).

Arthurs and Kelly consider an initial apparatus state
with wave function of the form
\begin{equation*}
   \overlap{\Mx, \Mp}{\pt} 
=  \frac{2}{\sqrt{h}} e^{ - \frac{1}{\lambda^2} \Mx^2 - \frac{\lambda^2}{\hbar^2} \Mp^2}
\end{equation*}
The reader may easily verify, that for this choice of $\ket{\pt}$ the errors are given by
\begin{equation*}
\begin{aligned}
   \RErr x & =  & \PErr x & = \frac{\lambda}{\sqrt{2}} \\
   \RErr p & =  & \PErr p & = \frac{\hbar}{\sqrt{2} \lambda}
\end{aligned}
\end{equation*}
We see, that the apparatus states considered by Arthurs and Kelly
minimise, both the product $\RErr x \, \RErr p$, and the product $\PErr x \, \PErr p$.
In other words, they maximise both the retrodictive and the predictive accuracy of the
measurement.
\section{Unbiased Measurements}
\label{sec:  UnbiasMeas}
After introducing the particular process which we discussed in the last
section, Arthurs and Kelly~(1965) go on to define a general class of measurement
processes.  They show that their
extended uncertainty principle, relation~(\ref{eq:  AKExtUncertPrinc}) above, holds for every
process in this class (also see Arthurs and Goodman, 1988).  It is natural to ask
whether the error-error and error-disturbance 
relations~(\ref{eq:  AKRetErrRel}--\ref{eq:  AKErrDisRels})  also generalise.

As before, the system is assumed to interact with a 
measuring apparatus, characterised by two pointer observables
$\MxOp$, $\MpOp$ which commute with each other, and with the observables
being measured $\hat{x}$, $\hat{p}$.  However, the apparatus may now have
additional degrees of freedom, apart from these two.  

Let  $\hat{U}$ be the 
unitary evolution
operator describing the measurement interaction, and define  
error and disturbance operators as in the last section.  Arthurs and Kelly
assume that the evolution operator $\hat{U}$ and initial apparatus state
$\ket{\pt}$ are such that
\begin{equation}
\begin{split}
        \bmat{\psi \otimes \pt}{\MxfOp}{\psi \otimes \pt} 
  &  =  \bmat{\psi \otimes \pt}{\xiOp}{\psi \otimes \pt}
\\
        \bmat{\psi \otimes \pt}{\MpfOp}{\psi \otimes \pt} 
  &  =  \bmat{\psi \otimes \pt}{\piOp}{\psi \otimes \pt}
\end{split}
\label{eq:  RetUnbiasMeanPt}
\end{equation}
uniformly, for every initial system state $\ket{\psi}$.  In our terminology
this condition amounts to the requirement that there be no systematic errors
of retrodiction:
\begin{equation}
\begin{split}
        \bmat{\psi \otimes \pt}{\Exi}{\psi \otimes \pt} 
  &  =  0
\\
        \bmat{\psi \otimes \pt}{\Epi}{\psi \otimes \pt} 
  &  =  0
\end{split}
\label{eq:  RetUnbiasMeanErr}
\end{equation}
for all  $\ket{\psi}$.  We will accordingly
refer to such a measurement as \emph{retrodictively unbiased}.

It is natural also to impose the requirement that the measurement
be \emph{predictively unbiased}:
\begin{equation*}
\begin{split}
        \bmat{\psi \otimes \pt}{\Exf}{\psi \otimes \pt} 
  &  =  0
\\
        \bmat{\psi \otimes \pt}{\Epf}{\psi \otimes \pt} 
  &  =  0
\end{split}
\end{equation*}
for all  $\ket{\psi}$.  

We now show, that all six
of the error-error and error-disturbance 
relations~(\ref{eq:  AKRetErrRel}--\ref{eq:  AKErrDisRels}) continue
to hold for every measurement which is both retrodictively and predictively
unbiased.  We will do so by using  a method similar to the one used by
Arthurs and Kelly to prove their extended uncertainty
principle~(\ref{eq:  AKExtUncertPrinc}).

We begin with the predictive error relationship.  We have
\begin{equation*}
    \bcomm{\xfOp}{\pfOp} = \hat{U}^{\dagger}\bcomm{\hat{x}}{\hat{p}}\hat{U} = i \hbar
\end{equation*}
This is the only non-vanishing commutator between members of the set
$\xfOp$, $\pfOp$, $\MxfOp$, $\MpfOp$.  Therefore
\begin{equation*}
   \bcomm{\Exf}{\Epf} 
=  \bcomm{\left(\MxfOp - \xfOp\right)}{\left( \MpfOp - \pfOp \right)} 
=  i \hbar 
\end{equation*}
We deduce
\begin{equation*}
   \PErr x \, \PErr p  \ge \frac{\hbar}{2}
\end{equation*}
We made no use of the assumption that the measurement is unbiased in deriving 
this inequality.  The predictive error relationship therefore holds quite generally.
The remaining relationships mix Heisenberg picture operators defined at different times,
and for these we must work a little harder.

Given an initial system state $\ket{\psi}$, let $\ket{\psi'} = \xiOp \ket{\psi}$.
If the measurement is retrodictively unbiased we then have, from the proposition
proved in the Appendix,
\begin{equation}
    \bmat{\psi \otimes \pt}{\Exi \xiOp}{\psi \otimes \pt}
=   \bmat{\psi \otimes \pt}{\Exi }{\psi' \otimes \pt}
=   0
\label{eq:  RetUnbiasEXrel}
\end{equation}
Similarly
\begin{equation}
    \bmat{\psi \otimes \pt}{ \xiOp \Exi }{\psi \otimes \pt} = 0
\label{eq:  RetUnbiasXErel}
\end{equation}
and
\begin{align*}
     \bmat{\psi \otimes \pt}{\Exi  \piOp }{\psi \otimes \pt}
&  =  \bmat{\psi \otimes \pt}{ \piOp \Exi }{\psi \otimes \pt} 
  =  0
\\
     \bmat{\psi \otimes \pt}{ \Epi \xiOp }{\psi \otimes \pt}
&  =  \bmat{\psi \otimes \pt}{ \xiOp \Epi }{\psi \otimes \pt} 
  =  0
\\
    \bmat{\psi \otimes \pt}{ \Epi \piOp }{\psi \otimes \pt}
&  = \bmat{\psi \otimes \pt}{ \piOp \Epi  }{\psi \otimes \pt} 
  = 0
\end{align*}
Using these equations, and the definitions of $\Exi$, $\Epi$ it is readily inferred
\begin{equation}
\begin{aligned}
    \bmat{\psi \otimes \pt}{\comm{\xiOp}{\MxfOp}}{\psi \otimes \pt} & = 0  \hspace{0.3 in}
  & \bmat{\psi \otimes \pt}{\comm{\MxfOp}{\piOp}}{\psi \otimes \pt} & = i \hbar
\\
    \bmat{\psi \otimes \pt}{\comm{\xiOp}{\MpfOp}}{\psi \otimes \pt} & = i \hbar  \hspace{0.3 in}
  & \bmat{\psi \otimes \pt}{\comm{\MpfOp}{\piOp}}{\psi \otimes \pt} & = 0
\end{aligned}
\label{eq:  RetUnbiasComRels}
\end{equation}
which, together with the fact that $\MxfOp$ and $\MpfOp$ commute, implies
\begin{equation*}
    \bmat{\psi \otimes \pt}{\comm{\Exi}{\Epi}}{\psi \otimes \pt}
=   \bmat{   \psi \otimes \pt  
        }{   \comm{\left(\MxfOp - \xiOp\right)}{\left(\MpfOp - \piOp\right)}
        }{   \psi \otimes \pt
        }
=   - i \hbar
\end{equation*}
for all $\ket{\psi}$.   Consequently
\begin{equation}
    \RErr x \, \RErr p  \ge \frac{\hbar}{2}
\label{eq:  RetErrRelB}
\end{equation}
In proving this inequality we only used the assumption that the measurement is
retrodictively unbiased.  The retrodictive error relationship is therefore valid under the
same set of assumptions which Arthurs and Kelly make in order to prove their
extended uncertainty principle.

Suppose, now, that the measurement is both retrodictively and 
predictively unbiased.  Then, by an
argument similar to that used in proving 
equations~(\ref{eq:  RetUnbiasComRels}), we find
\begin{equation*}
\begin{aligned}
    \bmat{\psi \otimes \pt}{\comm{\xiOp}{\xfOp}}{\psi \otimes \pt} & = 0  \hspace{0.3 in}
  & \bmat{\psi \otimes \pt}{\comm{\xfOp}{\piOp}}{\psi \otimes \pt} & = i \hbar
\\
    \bmat{\psi \otimes \pt}{\comm{\xiOp}{\pfOp}}{\psi \otimes \pt} & = i \hbar  \hspace{0.3 in}
  & \bmat{\psi \otimes \pt}{\comm{\pfOp}{\piOp}}{\psi \otimes \pt} & = 0
\end{aligned}
\end{equation*}
Therefore
\begin{equation*}
    \bmat{\psi \otimes \pt}{\bcomm{\Exi}{\Dp}}{\psi \otimes \pt}
=   \bmat{   \psi \otimes \pt  
        }{   \bcomm{\left(\MxfOp - \xiOp\right)}{\left(\pfOp - \piOp\right)}
        }{   \psi \otimes \pt
        }
=   - i \hbar
\end{equation*}
Similarly
\begin{equation*}
    \bmat{\psi \otimes \pt}{\bcomm{\Exf}{\Dp}}{\psi \otimes \pt} = - i \hbar
\end{equation*}
and
\begin{align*}
    \bmat{\psi \otimes \pt}{\bcomm{\Epi}{\Dx}}{\psi \otimes \pt} & = i \hbar  \\
    \bmat{\psi \otimes \pt}{\bcomm{\Epf}{\Dx}}{\psi \otimes \pt} & =  i \hbar
\end{align*}
Hence
\begin{equation}
\begin{aligned}
  \RErr x \, \Dist p & \ge \frac{\hbar}{2} \hspace{0.5 in} &
  \RErr p \, \Dist x & \ge \frac{\hbar}{2}
\\
  \PErr x \, \Dist p & \ge \frac{\hbar}{2} \hspace{0.5 in} &
  \PErr p \, \Dist x & \ge \frac{\hbar}{2}
\end{aligned}
\label{eq:  ErrDistRelsB}
\end{equation}
It would be interesting to see if one can remove the restriction to measurement
processes which are retrodictively unbiased [in the case of inequality
(\ref{eq:  RetErrRelB})], or retrodictively and predictively unbiased  [in the case of 
inequalities
(\ref{eq:  ErrDistRelsB})].
\section{The Arthurs-Kelly Principle and Related Inequalities}
For the sake of completeness we briefly indicate the connection between
the inequalities proved in the last section, and the extended uncertainty principle
of Arthurs and Kelly (1965).

Suppose that the measurement is retrodictively unbiased.  Then
\begin{equation*}
        \bmat{\psi \otimes \pt}{\MxfOp}{\psi \otimes \pt} 
     =  \bmat{\psi \otimes \pt}{\xiOp}{\psi \otimes \pt}
\end{equation*}
In view of equations~(\ref{eq:  RetUnbiasEXrel}) 
and~(\ref{eq:  RetUnbiasXErel}) we also have
\begin{align*}
        \bmat{\psi \otimes \pt}{\MxfOp^2}{\psi \otimes \pt} 
&    =  \bmat{\psi \otimes \pt}{\left( \xiOp + \Exi\right)^2}{\psi \otimes \pt}
\\
&    =  \bmat{\psi \otimes \pt}{\xiOp^2}{\psi \otimes \pt}
         + \bmat{\psi \otimes \pt}{\Exi^2}{\psi \otimes \pt}
\end{align*}
Using equations~(\ref{eq:  RetUnbiasMeanPt}) and~(\ref{eq:  RetUnbiasMeanErr})
we deduce
\begin{equation}
    \left( \Delta \Mxf \right)^2 
=   \left( \Delta x_{\mathrm{i}} \right)^2 + \left( \RErr x \right)^2
\label{eq:  XPointUncert}
\end{equation}
where $\Delta \Mxf$, $\Delta x_{\mathrm{i}}$ represent uncertainties calculated in the
usual way, according to the prescription of 
equation~(\ref{eq:  UncertsDef}).
Similarly
\begin{equation}
    \left( \Delta \Mpf \right)^2 
=   \left( \Delta p_{\mathrm{i}} \right)^2 + \left( \RErr p \right)^2
\label{eq:  PPointUncert}
\end{equation}
We see that $\RErr x$ and $\RErr p$ determine the increases in the variances of the
distribution of results, as compared with the intrinsic variances of the 
initial system state.

Equations~(\ref{eq:  XPointUncert}) 
and~(\ref{eq:  PPointUncert}), together with Kennard's 
inequality~(\ref{eq:  UncertRel}) and the retrodictive error 
relationship~(\ref{eq:  RetErrRelB}), imply
\begin{align}
    \left( \Delta \Mxf \right)^2  \, \left( \Delta \Mpf \right)^2 
& = \left(  \left( \Delta x_{\mathrm{i}} \right)^2 + \left( \RErr x \right)^2
    \right)
    \left(  \left( \Delta p_{\mathrm{i}} \right)^2 + \left( \RErr p \right)^2
    \right)
\nonumber
\\
& \ge \frac{\hbar^2}{4}
      \left(  \left( \Delta x_{\mathrm{i}} \right)^2 + \left( \RErr x \right)^2
      \right)
      \left(  \frac{1}{ \left( \Delta x_{\mathrm{i}} \right)^2  }
             + \frac{1}{    \left( \RErr x \right)^2   }
      \right)
\nonumber
\\
& =   \frac{\hbar^2}{4}
      \left( 2 + \frac{  \left( \Delta x_{\mathrm{i}} \right)^2 
                     }{  \left( \RErr x \right)^2  
                      }
              +  \frac{  \left( \RErr x \right)^2 
                     }{  \left( \Delta x_{\mathrm{i}} \right)^2 
                      }
      \right)
\nonumber
\\
& \ge \hbar^2
\label{eq:  AKExtUncertPrincB}
\end{align}
which is the extended principle of Arthurs and Kelly.

If the measurement is both retrodictively and predictively unbiased we can 
also prove, by a similar argument,
\begin{equation}
\begin{split}
    \left( \Delta x_{\mathrm{f}} \right)^2 
& = \left( \Delta x_{\mathrm{i}} \right)^2 + \left( \Dist x \right)^2
\\
    \left( \Delta p_{\mathrm{f}} \right)^2 
& = \left( \Delta p_{\mathrm{i}} \right)^2 + \left( \Dist p \right)^2
\end{split}
\label{eq:  FinSteUncerts}
\end{equation}
showing how the mean square disturbances determine the extent of the increase in the 
system state variances.  These inequalities do not imply an increase in the lower
bound on the product 
$\Delta x_{\mathrm{f}} \, \Delta p_{\mathrm{f}}$, because the disturbances can
both be made arbitrarily small (at the expense of making the measurement very inaccurate).
The lower bound for the final system state uncertainties is therefore the same as that
for the initial state ones:  namely,
\begin{equation*}
  \Delta x_{\mathrm{f}} \, \Delta p_{\mathrm{f}}  \ge \frac{\hbar}{2}
\end{equation*}
On the other hand, the lower bound on the products $\Delta x_{\mathrm{f}} \, \Delta \Mpf$
and $\Delta \Mxf \, \Delta p_{\mathrm{f}}$ is larger than the one set by
Kennard's inequality.  In fact, (\ref{eq:  XPointUncert}), 
(\ref{eq:  PPointUncert}) and (\ref{eq:  FinSteUncerts}),
together with the error-disturbance relations~(\ref{eq:  ErrDistRelsB}) are readily seen
to imply
\begin{equation*}
\begin{split}
   \Delta x_{\mathrm{f}} \, \Delta \Mpf  & \ge \hbar \\
   \Delta \Mxf \, \Delta p_{\mathrm{f}}  & \ge \hbar
\end{split}
\end{equation*}
\section{The Question of Interpretation}
We now come to the question which we have been ignoring up to now.  We have been referring
to the quantities $\RErr x$, $\RErr p$, $\PErr x$, $\PErr p$ as experimental
errors, and the quantities $\Dist x$, $\Dist p$ as disturbances.  
Is this terminology really justified?

Let us begin with the quantities $\PErr x$, $\PErr p$.   The observables
$\MxfOp$, $\MpfOp$, $\xfOp$ commute, and can therefore be
simultaneously determined with arbitrarily high precision.  Alternatively, one may determine
the values of $\MxfOp$, $\MpfOp$ without perturbing  $\xfOp$.  We may therefore envisage
a procedure, in which one \emph{first} makes a highly accurate determination of
the meter readings, and then  
checks the value of $\MxfOp$ by making an (immediately) \emph{subsequent} 
highly accurate determination of 
$\xfOp$.  Suppose that one takes numerous copies of the system, all prepared in the
same state, performs this procedure on each of them, and calculates the rms value
of the differences $\Mxf - x_{\mathrm{f}}$.  Then, provided that the
verification of $\xfOp$ is carried out immediately after the determination of 
$\MxfOp$, $\MpfOp$, the quantity which results will almost certainly be no 
larger than an amount
$\sim \PErr x$.

We can equally well envisage a procedure in which one makes a second, verificatory
measurement of $\pfOp$ immediately after recording the meter readings.  If one repeated this
procedure many times then the rms value of the differences $\Mpf - p_{\mathrm{f}}$
would almost certainly be no larger than an amount $\sim \PErr p$.

Suppose, now, that one has recorded the meter readings to be $\Mxf$, $\Mpf$.  What can be
deduced about the likely state of the system?  It is, of course, impossible to check
the values of both $\xfOp$ and $\pfOp$ to arbitrarily high precision.  It is, however,
possible to counterfactually say, that if one were to make a single, 
immediately subsequent high
precision measurement of $\xfOp$, then the result would typically differ from $\Mxf$
by an amount $\sim \PErr x$.  It is also possible to counterfactually say, that 
if one were to make a single, immediately subsequent high precision 
measurement of $\pfOp$, then
the result would typically differ from $\Mpf$ by an amount 
$\sim \PErr p$.  There is therefore a well-defined sense in which it 
may justifiably be said, that
in recording the meter readings $\Mxf$, $\Mpf$, one has simultaneously determined
the final values of the position and momentum of the system to accuracy
$\pm \PErr x$ and $\pm \PErr p$ respectively.  

The interpretation of the quantities 
$\RErr x$, $\RErr p$ is less straightforward.  This is because the
observables $\xiOp$ and $\MpfOp$ do not commute.  Nor do the observables
$\piOp$ and $\MxfOp$  (see equations~(\ref{eq:  RetUnbiasComRels}) in the last 
section).  This means that the act of making a  precise
determination of the meter readings $\MxfOp$, $\MpfOp$ precludes one from 
making a precise determination of the values of either $\xiOp$ or 
$\piOp$.  It follows, that in the case
of the quantities $\RErr x$, $\RErr p$, we cannot carry through an analysis analogous to the
one given in the preceding paragraphs for 
$\PErr x$, $\PErr p$.

There is an obvious physical reason why one might expect the concept of retrodictive error
to be more problematic than the concept of predictive error.  
The effect of carrying out a measurement, and
recording the meter readings, is (as we have seen) to put the 
system into a state such that its \emph{final} position and momentum are confined, with high
probability, to a localised region of phase space.  However, this is an effect
produced by the measurement process itself.  If the uncertainties of the 
initial system state are large, then the \emph{initial} values of the position and momentum will
be quite indeterminate.  In such a case the concept of retrodictive error does not really
make sense.  At least, the concept does not make sense if it is defined in anything like the 
classical manner.

Classically, one thinks of the retrodictive error as the difference between the 
result of the measurement, and the value which the quantity being measured 
did take, before the measurement was carried out.  In quantum mechanics, however,
the quantity being measured may not have had a well-defined initial value.

Nevertheless, there is at least one situation in which it is possible to attach a meaning 
to the concept that is similar to the meaning which it has in classical physics.
In section~\ref{sec:  intro} we stated that one of the reasons that an 
error principle is needed is to justify the assumption (which plays an essential
role in experimental physics) that it is normally possible to determine both the
position and momentum of a macroscopic object to within a very small percentage error.
Suppose that it is a measurement such as this which is in question.  Then it will usually
be reasonable to assume that the initial system state is a localised wave packet.  In other words,
the uncertainties $\Delta x_{\mathrm{i}}$, $\Delta p_{\mathrm{i}}$ may be assumed to be 
small.  The purpose of the measurement is to determine the mean values
$\overline{x}_{\mathrm{i}} = \mat{\psi}{\xiOp}{\psi}$ and
$\overline{p}_{\mathrm{i}} = \mat{\psi}{\piOp}{\psi}$.
If the measurement is retrodictively unbiased
\begin{equation*}
\begin{split}
   \bmat{\psi \otimes \pt}{\MxfOp }{\psi \otimes \pt} & = \overline{x}_{\mathrm{i}} \\
   \bmat{\psi \otimes \pt}{\MpfOp }{\psi \otimes \pt} & = \overline{p}_{\mathrm{i}}
\end{split}
\end{equation*}
In view of equations~(\ref{eq:  XPointUncert}) and~(\ref{eq:  PPointUncert}) we then
have
\begin{equation}
\begin{split}
   \bmat{\psi \otimes \pt}{\left(\MxfOp - \overline{x}_{\mathrm{i}}\right)^2}{\psi \otimes \pt}
& = \left( \Delta \Mxf \right)^2
  = \left( \Delta x_{\mathrm{i}}\right)^2 + \left( \RErr x \right)^2
\\
   \bmat{\psi \otimes \pt}{\left(\MpfOp - \overline{p}_{\mathrm{i}}\right)^2}{\psi \otimes \pt}
& = \left( \Delta \Mpf \right)^2
  = \left( \Delta p_{\mathrm{i}}\right)^2 + \left( \RErr p \right)^2
\end{split}
\label{eq:  LocWpcktRErrs}
\end{equation}
It follows, that the process determines the values of $\overline{x}_{\mathrm{i}}$,
$\overline{p}_{\mathrm{i}}$ up to an uncertainty of 
$\pm \sqrt{\left( \Delta x_{\mathrm{i}}\right)^2 + \left( \RErr x \right)^2}$
in the determination of $\overline{x}_{\mathrm{i}}$,
and  $\pm \sqrt{\left( \Delta p_{\mathrm{i}}\right)^2 + \left( \RErr p \right)^2}$
in the determination of $\overline{p}_{\mathrm{i}}$.  The quantities
$\RErr x$, $\RErr p$ represent the part of the total error which arises from the measurement
process itself, as opposed to the intrinsic uncertainties of the initial state.
In other words, they represent the experimental errors.

If the initial system state is not a localised wave packet then the classical,
or ordinary intuitive concept of retrodictive error does not apply.
One should realise, however, that this has nothing specially to do with the fact that
we are considering simultaneous measurements of position and momentum.
Exactly the same problem arises when interpreting the quantity
$\Delta x_{\mathrm{measurement}}$ defined by Braginsky and Khalili (1992)
for single measurements of position only.  It is a simple consequence
of the fact that quantum mechanical observables need not take determinate 
values.  This feature of the quantum mechanical theory of measurement is 
sometimes expressed by saying, that we create the value by the act of measuring it.

Although they are then not  interpretable as errors in the classical sense, the
quantities $\RErr x$, $\RErr p$ are still defined when the initial system
state does not take the form of a localised wave packet.  Furthermore, they 
still play a role in characterising the ``goodness'', or ``faithfulness'' of 
the measurement.  Suppose, for instance, that the initial system state is a 
superposition of a finite or countable number of well-separated,
localised wave packets:
\begin{equation*}
   \ket{\psi}  =  \sum_{n} c_{n} \ket{\chi_n}
\end{equation*}
In this expression $\ket{\psi}$ and the $\ket{\chi_n}$ are all assumed
to be normalised.  Define
\begin{equation*}
\begin{split}
   \overline{x}_{\mathrm{i} n} & = \mat{\chi_n}{\hat{x}}{\chi_n} \\
   \overline{p}_{\mathrm{i} n} & = \mat{\chi_n}{\hat{p}}{\chi_n}
\end{split}
\end{equation*}
and
\begin{equation*}
\begin{split}
   l_{\mathrm{X}} & = \min_{n \neq m} |\overline{x}_{\mathrm{i} n} -\overline{x}_{\mathrm{i} m}| \\
   l_{\mathrm{P}} & = \min_{n \neq m} |\overline{p}_{\mathrm{i} n} -\overline{p}_{\mathrm{i} m}|
\end{split}
\end{equation*}
For the sake of simplicity assume that the states $\ket{\chi_n}$ all have the same 
intrinsic uncertainties $\sigma_{\mathrm{X}}$, $\sigma_{\mathrm{P}}$:
\begin{equation*}
\begin{split}
       \sigma_{\mathrm{X}} 
   & = \sqrt{\bmat{\chi_n}{\left(\hat{x} - \overline{x}_{\mathrm{i} n}\right)^2}{\chi_n}}\\
       \sigma_{\mathrm{P}} 
   & = \sqrt{\bmat{\chi_n}{\left(\hat{p} - \overline{p}_{\mathrm{i} n}\right)^2}{\chi_n}}\\
\end{split}
\end{equation*}
for all $n$.  The assumption that the wave packets are well-separated means that
$\sigma_{\mathrm{X}} \ll l_{\mathrm{X}}$ and
$\sigma_{\mathrm{P}} \ll l_{\mathrm{P}}$.  We then have
\begin{equation*}
   \overlap{\chi_{n}}{\chi_{m}} \approx \delta_{n m}
\end{equation*}
and
\begin{equation*}
   \sum_{n} | c_{n} |^2 \approx 1
\end{equation*}
Now surround each point $\left( \overline{x}_{\mathrm{i} n},\overline{p}_{\mathrm{i} n} \right)$
with a region $\mathscr{R}_n$ whose dimensions are large compared with 
the intrinsic uncertainties $\sigma_{\mathrm{X}}$, $\sigma_{\mathrm{P}}$, but
small compared with the minimum separations $l_{\mathrm{X}}$, $l_{\mathrm{P}}$:
\begin{equation*}
  \mathscr{R}_n 
= \left \{ (x,p) \in \mathbb{R}^2 \colon 
          \left| x - \overline{x}_{\mathrm{i} n} \right| < d_{\mathrm{X}}, \;
          \left| p - \overline{p}_{\mathrm{i} n} \right| < d_{\mathrm{P}}    
  \right \}
\end{equation*}
where $\sigma_{\mathrm{X}} \ll d_{\mathrm{X}} \ll l_{\mathrm{X}}$ and
$\sigma_{\mathrm{P}} \ll d_{\mathrm{P}} \ll l_{\mathrm{P}}$.
Suppose that we also have $\RErr x \ll d_{\mathrm{X}}$ and 
$\RErr p \ll d_{\mathrm{P}}$.  In view of~(\ref{eq:  LocWpcktRErrs})
the function 
$\bigl| \bmat{x,\mu_{\mathrm{X}},\mu_{\mathrm{P}}}{\hat{U}}{\chi_{m} \otimes \pt} \bigr|^2$
is then concentrated on the set $\mathbb{R} \times \mathscr{R}_m$.  Hence
\begin{equation*}
   \int_{\mathbb{R} \times \mathscr{R}_{n} }  d x d \mu_{\mathrm{X}} d \mu_{\mathrm{P}} \,
      \bigl| \bmat{x,\mu_{\mathrm{X}},\mu_{\mathrm{P}}}{\hat{U}}{\chi_{m} \otimes \pt} \bigr|^2
\approx \delta_{n m}
\end{equation*}
Consequently
\begin{equation*}
  \int_{\mathbb{R} \times \mathscr{R}_{n}  }  d x d \mu_{\mathrm{X}} d \mu_{\mathrm{P}} \,
    \bigl| \bmat{x,\mu_{\mathrm{X}},\mu_{\mathrm{P}}}{\hat{U}}{\psi \otimes \pt} \bigr|^2
\approx \left| c_{n} \right|^2
\label{eq:  InterpRetErrB}
\end{equation*}
In words:  the probability that the final pointer positions will be in 
the vicinity of the 
point $\left( \overline{x}_{\mathrm{i} n},\overline{p}_{\mathrm{i} n} \right)$
is approximately $\left| c_{n} \right|^2$, provided that the rms errors of 
retrodiction are sufficiently small.

This result may be regarded as a generalisation 
of the following well-known fact regarding measurements
of a single, discrete observable $\hat{A}$.  Let $\ket{a}$ be the eigenstate
of $\hat{A}$ with eigenvalue $a$, and suppose that the system is in the state
\begin{equation*}
    \ket{\psi} = \sum_{a} c_{a} \ket{a}
\end{equation*}
Suppose that one performs a perfectly precise measurement of $\hat{A}$.  Then the
probability of recording the value $a$ is $|c_{a}|^2$.  The analogy between this proposition
and the result just proved lends some support to the suggestion, that processes
of the kind described by Arthurs and Kelly may be regarded as simultaneous measurements
of non-commuting observables.

Finally, let us consider the interpretation of the quantities 
$\Dist x$, $\Dist p$.   
Suppose that the measurement is both retrodictively
and predictively unbiased.  By an argument similar to the one leading
to equations~(\ref{eq:  LocWpcktRErrs}) we find
\begin{equation*}
\begin{split}
    \bmat{\psi \otimes \pt}{\left(\xfOp - \overline{x}_{\mathrm{i}}\right)^2}{\psi \otimes \pt}
& = \left( \Delta x_{\mathrm{f}} \right)^2
  = \left( \Delta x_{\mathrm{i}} \right)^2 + \left( \Dist x \right)^2 
\\
    \bmat{\psi \otimes \pt}{\left(\pfOp - \overline{p}_{\mathrm{i}}\right)^2}{\psi \otimes \pt}
& = \left( \Delta p_{\mathrm{f}} \right)^2
  = \left( \Delta p_{\mathrm{i}} \right)^2 + \left( \Dist p \right)^2
\end{split}
\label{eq:  FinSteUncertsB}
\end{equation*}
where $\overline{x}_{\mathrm{i}}$, $\overline{p}_{\mathrm{i}}$ are the 
expectation values of $\xiOp$, $\piOp$, as before.  The effect of the measurement
process on the system state is to leave the expectation value of 
$\hat{x}$ (respectively $\hat{p}$) unchanged, while increasing the variance by 
an amount $\left(\Dist x \right)^2$ (respectively, $\left(\Dist p\right)^2$).  There is thus
a well-defined sense 
in which the quantities 
$\Dist x$, $\Dist p$ provide a numerical indication of the extent to 
which the measurement disturbs the state of the system.

\section{Conclusion}
Does quantum mechanics allow 
for the existence of simultaneous measurements of 
position and momentum?  For ourselves we can see 
no clear objection to the use of the term ``measurement'' to refer to the kind of process described
by Arthurs and Kelly.  
However, it must be admitted, that in so far as the question at issue is one of nomenclature,
it probably does not have a once-and-for-all right answer.  Such questions are,
in the end, a matter of taste.

What is not a matter of taste is the fact that processes of the kind considered are of
some importance in the field of quantum optics.  This is true irrespective of the name by
which one chooses to describe them.  
If the quantities introduced in this paper are to be of any interest they must be
justified in the same way, in terms of their usefulness.
Braginsky and Khalili have shown that the relationship they derive is a useful
tool in the analysis of single measurements of $\hat{x}$ or $\hat{p}$ separately.   It 
seems not unreasonable to suppose that the relationships derived in this paper may be no less
useful in the analysis of simultaneous measurements of $\hat{x}$ and $\hat{p}$ together.
At the least, they seem worthy of further investigation.
\section*{Appendix}
In section~\ref{sec:  UnbiasMeas} we rely on a proposition which forms the basis of the argument
in both Arthurs and Kelly~(1965) and Arthurs and Goodman~(1988).  However, in 
neither case do the authors actually prove this proposition.  Since it is not
entirely obvious we give the proof here.

\begin{proposition}
Let $\mathscr{H}_1$,  $\mathscr{H}_2$ be two Hilbert spaces
and let $\hat{A}$ be a (possibly unbounded) linear operator defined on
the product space $\mathscr{H}_1 \otimes \mathscr{H}_2$.  Let
$\mathscr{D} \subseteq \mathscr{H}_1 \otimes \mathscr{H}_2$ be the domain of
$\hat{A}$.  Let 
$\ket{\phi}$ be a fixed vector $\in \mathscr{H}_2$.  Suppose that
$\mathscr{H}_1 \otimes \ket{\phi} \subseteq \mathscr{D}$, and suppose also that
\begin{equation}
    \bmat{\psi \otimes \phi}{ \hat{A} }{\psi \otimes \phi} = 0
\label{eq:  AppPropCond}
\end{equation}
for all $\ket{\psi} \in \mathscr{H}_1$.  Then 
\begin{equation*}
    \bmat{\psi \otimes \phi}{ \hat{A} }{\psi' \otimes \phi} = 0
\end{equation*}
for all $\ket{\psi}$, $\ket{\psi'} \in \mathscr{H}_1$
\end{proposition}
\begin{proof}
The result is proved in essentially the same way as
(for example) Proposition~2.4.3 in Kadison and Ringrose (1983).  Given arbitrary
$\ket{\psi}$, $\ket{\psi'} \in \mathscr{H}_1$ we have the identity
\begin{align}
&    \bmat{\psi \otimes \phi}{\hat{A}}{\psi' \otimes \phi}
\nonumber
\\
& \hspace{0.2 in }
=  \frac{1}{4} \left( \bmat{(\psi + \psi') \otimes \phi}{\hat{A}}{(\psi+\psi') \otimes \phi}
                       - \bmat{(\psi - \psi') \otimes \phi}{\hat{A}}{(\psi-\psi') \otimes \phi}
                 \right.
\nonumber
\\
& \hspace{0.5 in}
      \left.   - i \bmat{(\psi + i \psi') \otimes \phi}{\hat{A}}{(\psi + i \psi') \otimes \phi}
               + i \bmat{(\psi - i \psi') \otimes \phi}{\hat{A}}{(\psi - i \psi') \otimes \phi}
      \right)
\nonumber
\end{align}
Using equation~(\ref{eq:  AppPropCond}) we deduce
\begin{equation*}
    \bmat{\psi \otimes \phi}{\hat{A}}{\psi' \otimes \phi} = 0
\end{equation*}

\end{proof}
\section*{References}
\begin{enumerate}
\item
  Arthurs, E., and Goodman, M.S. (1988).
   \emph{Physical Review Letters}, \textbf{60}, 2447.
\item
		Arthurs, E., and Kelly, J.L. Jr. (1965).
			\emph{Bell System  Technical  Journal}, \textbf{44}, 725.
\item
  Ballentine, L.F.~(1970).
  \emph{Reviews of Modern Physics}, \textbf{42}, 358.
\item
  Bohm, D. (1951).
    \emph{Quantum Theory},
    Prentice Hall, New York.
\item
  Braginsky, V.B., and Khalili, F.\ Ya (1992).
  \emph{Quantum Measurement}, \linebreak K.S.~Thorne, ed., Cambridge University Press, Cambridge.
\item
  Braunstein, S.L., Caves, C.M., and Milburn G.J. (1991).  
  \emph{Physical Review A}, \textbf{43}, 1153.
\item
  Busch, P., and Lahti, P.J. (1984).
  \emph{Physical Review D}, \textbf{29}, 1634.
\item
  de Muynck, W.M., De Baere, W., and Martens, H. (1994).
   \emph{Foundations of Physics}, \textbf{24}, 1589.
\item
  	Heisenberg, W. (1927).
  		\emph{Zeitschrift f\"{u}r Physik}, \textbf{43}, 172.
    Reprinted in 
    \emph{Quantum Theory and Measurement},  
    J.A.~Wheeler and W.H.~Zurek, eds.,
    Princeton University Press, Princeton N.J., 1983.
\item
		Heisenberg, W. (1930).
    \emph{The Physical Principles of the Quantum Theory},
     C.~Eckart and F.C.~Hoyt, trans., University of Chicago Press, Chicago, 1930; 
    Dover Publications, New York, 1949.
\item
  Hilgevoord, J., and Uffink, J. (1990).
    In \emph{Sixty-Two Years of Uncertainty},  A.I.~Miller, ed., Plenum Press, New York.
\item
  Holevo, A.S. (1982).  \emph{Probabilistic and Statistical Aspects of Quantum  Theory},
  North-Holland, Amsterdam.
\item
  Kadison, R.V., and Ringrose, J.R. (1983).  
  \emph{Fundamentals of the Theory of Operator Algebras}, Academic Press,
  New York.
\item
	Kennard, E.H. (1927).
  		\emph{Zeitschrift f\"{u}r Physik}, \textbf{44}, 326.
\item
 Leonhardt, U., and Paul, H. (1993)
 \emph{Journal of Modern Optics}, \textbf{40}, 1745.
\item
  Leonhardt, U., B\"{o}hmer, B. and Paul, H. (1995).
   \emph{Optics Communications}, \textbf{119}, 296.
\item
  Martens, H., and de Muynck, W.M. (1992).
  \emph{Journal of Physics A}, \textbf{25}, 4887.
\item
  Prugove\v{c}ki, E. (1973).
  \emph{Foundations of Physics}, \textbf{3}, 3.
\item
  Prugove\v{c}ki, E. (1975).
  \emph{Foundations of Physics}, \textbf{5}, 557.
\item
  Prugove\v{c}ki, E. (1984).
   \emph{Stochastic Quantum Mechanics and Quantum Spacetime}, Reidel, Dordrecht.
\item
  Raymer, M.G. (1994).
    \emph{American  Journal of Physics}, \textbf{62}, 986.
\item
  Stenholm, S. (1992).
  \emph{Annals of Physics (N.Y.)}, \textbf{218}, 233.
\item
  T\"{o}rma, P., Stenholm S., and Jex, I. (1995).
  \emph{Physical Review A}, \textbf{52}, 4812.
\item
  Uffink, J. (1994).  
   \emph{International Journal of Theoretical Physics}, \textbf{33}, 199.
\item
  W\'{o}dkiewicz, K. (1987).
      \emph{Physics Letters A}, \textbf{124}, 207.
\end{enumerate}


\end{document}